\documentclass[twocolumn,showpacs,preprintnumbers,amsmath,amssymb]{revtex4}

\usepackage{graphicx}
\usepackage{dcolumn}
\usepackage{bm}

\newcommand*{\cdag}{$^{106}$Cd($\alpha,\gamma)^{110}$Sn }
\newcommand*{\cdan}{$^{106}$Cd($\alpha$,n)$^{109}$Sn }
\newcommand*{\cdap}{$^{106}$Cd($\alpha$,p)$^{109}$In }

\begin{document}

\preprint{APS/123-QED}

\title{Alpha-induced cross sections of $^{106}$Cd for the astrophysical p-process}

\author{Gy.\,Gy\"urky}%
 \email{gyurky@atomki.hu}
\author{G.G.\,Kiss}%
\author{Z.\,Elekes}%
\author{Zs.\,F\"ul\"op}%
\author{E.\,Somorjai}%
\affiliation{%
Institute of Nuclear Research (ATOMKI), H-4001 Debrecen, POB.51., Hungary}%
\author{A.\,Palumbo}%
\author{J.\,G\"orres}%
\author{H.Y.\,Lee}%
\author{W.\,Rapp}%
\author{M.\,Wiescher}%
\affiliation{%
University of Notre Dame, Notre Dame, Indiana 46556, USA}%
\author{N.\,\"Ozkan}%
\author{R.T.\,G\"uray}%
\author{G.\,Efe}%
\affiliation{%
Kocaeli University, Department of Physics, TR-41380
Umuttepe, Kocaeli, Turkey}%
\author{T.\,Rauscher}%
\affiliation{%
Universit\"at Basel, CH-4056 Basel, Switzerland}

\date{\today}

\begin{abstract}
The $^{106}$Cd($\alpha,\gamma)^{110}$Sn reaction cross section has been measured in the energy
range of the Gamow window for the astrophysical $p$-process
scenario. The cross sections for $^{106}$Cd($\alpha$,n)$^{109}$Sn and for $^{106}$Cd($\alpha$,p)$^{109}$In below the
($\alpha$,n) threshold have also been determined. The results are
compared with predictions of the statistical model code NON
SMOKER using different input parameters. The comparison shows
that a discrepancy for $^{106}$Cd($\alpha,\gamma)^{110}$Sn when using the standard optical potentials
can be removed with a different $\alpha$+$^{106}$Cd potential. Some
astrophysical implications are discussed.
\end{abstract}

\pacs{25.55.-e, 26.30.+k, 26.50.+x, 27.60.+j}%

\maketitle

\section{Introduction}

Stable heavy isotopes above iron ($Z>$\,26) can be classified into
three categories, $s$-, $r$- and $p$-nuclei corresponding to the
topology of the nuclide chart. The $s$-nuclei are located along the
valley of stability while the $r$- and $p$-nuclei can be found on the
neutron rich and proton rich side of the valley, respectively. The
names refer to the production process synthesizing the
corresponding isotopes. The $s$-process isotopes are produced by the
s (slow) neutron capture process in stellar helium and carbon
burning environments with steady neutron production through the
$^{13}$C, $^{17}$O, and $^{22}$Ne($\alpha$,n) reactions. The
$s$-process sites have been identified as low mass AGB stars ($M$ $<$
5 $M_{\odot}$) for the main $s$-process \cite{busso} and massive red
giant stars ($M$ $>$ 6 $M_{\odot}$) for the weak $s$-process
\cite{eleid}. On the other hand, the $r$-isotopes are produced by
the $r$ (rapid) neutron capture process which takes place in
explosive stellar environments providing a high neutron flux.
The $r$-process site is still under debate but the presently
favored candidates are type-II supernovae
\cite{qian} and merging neutron stars \cite{Freiburghaus99}. 
For the production of a number of isotopes located along the 
valley of stability both the $s$ and $r$ processes have their 
contributions.
The $p$-nuclei, however, cannot be produced by neutron capture
reactions. Their production mechanism, the $p$-process, has been
identified as a sequence of photodisintegration processes in a
high $\gamma$-flux scenario \cite{wh78}. The initial abundance distribution of
$s$- and $r$-nuclei at the $p$-process site is converted by subsequent
($\gamma$,n) reactions toward the neutron-deficient region. 
As the neutron threshold increases, competing ($\gamma,\alpha$) and
($\gamma,p$) photodisintegration processes branch the reaction
flow towards lower mass regions \cite{ra05}. The final $p$-nuclei abundance
distribution depends critically on the seed abundance distribution
as well as on the reaction flow which is determined by the
associated reaction rates and reaction branchings. A recent
detailed overview of the $p$-process and a discussion about possible
$p$-process sites can be found in \cite{ar03}.

The modeling of $p$-process nucleosynthesis requires a large
network of thousands of nuclear reactions involving stable and
unstable nuclei. The relevant astrophysical reaction rates which
are derived from the reaction cross sections are
necessary input to this reaction network. Their knowledge is
essential for $p$-process calculations. In some cases, the cross section of
$\gamma$-induced reactions can be measured directly by
photodissociation experiments \cite{mohr}; however, in such an 
experiment the target nucleus is always in its ground state while 
in stellar environments thermally populated excited states also 
contribute to the reaction rate. Thus theoretical considerations 
can not be avoided \cite{mo03}. Alternatively, the $\gamma$-induced reaction
cross sections can be calculated through "detailed 
balance" from the cross section of the inverse
capture reactions. While there are extensive compilations of
neutron capture data along the line of stability above the iron
region (e.g.\ \cite{bao}), there are only very few charged-particle cross sections
determined experimentally (despite substantial experimental
efforts in recent years). Therefore, the $p$-process rates involving
charged particles are still based mainly on
theoretical cross sections obtained from Hauser-Feshbach
statistical model calculations. It is particularly important to
study the charged particle photodissociation processes [($\gamma,\alpha$), ($\gamma$,p)] since those
determine the reaction flow towards the lower mass range.

Because of the large number of experimentally unaccessed or
unaccessible nuclei in astrophysical reaction networks,
statistical model codes in astrophysics focus on the prediction
of reaction rates from microscopic input or global parameterizations.
Contrary to standard Hauser-Feshbach calculations applied in other areas
of nuclear physics, they deliberately refrain from using local fine-tuning
by utilization of local nuclear properties. It is assumed that
such models allow better predictions for nuclei further off stability.
The trade-off is in a possible loss of accuracy locally while the average
deviation over a larger range of nuclei remains low. Nevertheless, also such
models have to be tested against local data in order to arrive at further
improvements.
There is an increasing number of (p,$\gamma$) reactions relevant
for the $p$-process which have been measured in recent years
\cite{la87,sa97,bo98,ch99,gyu01,ha01,oz01,ga03,gyu03,ts04}.
Generally, the statistical models are able to reproduce this
experimental data to better than a factor of two and the predictions
are not very dependent on the input parameters (e.g. optical
potentials). However, for ($\alpha,\gamma$) reactions only few
cases have been measured \cite{fu96,so98,oz01,ra01,oz06} or are 
under study \cite{ha05}. The experimental 
results show substantial discrepancies compared to
the model predictions. It has been suggested that these
discrepancies are related to insufficiencies in the $\alpha$-optical
potential. It is therefore important to measure ($\alpha,\gamma$)
cross sections at sub-Coulomb energies and compare the results
with the model calculations to identify the source of the observed
discrepancies.

In the present work the $\alpha$-capture cross section of
$^{106}$Cd is measured. This reaction is particularly important
since it focuses on the study of photodisintegration of a
$^{110}$Sn nucleus with $Z$\,=\,50 closed proton shell. Near closed
shells the level density is reduced and the statistical model may
not be fully applicable. The \cdag reaction is therefore a prime
example to test the validity of the Hauser-Feshbach 
approach in this mass region. In addition, while for 
higher $Z$ nuclei along the $p$-process path the
alpha threshold is negative, for $^{110}$Sn the $\alpha$-threshold
turns positive $S_{\alpha}$=1.136 MeV and increases towards lower
$Z$. This means that ($\gamma,\alpha$) photodissociation into the
alpha channel for even-even nuclei below $Z$\,=\,50 is reduced and the
reaction flow may become diverted towards the line of stability by
competing ($\gamma,p$) reactions \cite{ra05}. (The experimantal study of the 
(p,$\gamma$) reactions on $^{106}$Cd and $^{108}$Cd is in progress. Preliminary 
results are already available \cite {gyu06}). For nuclei between $N$\,=\,50 and $Z$\,=\,50
both proton and $\alpha$-photodissociation channels need to be
studied in detail to see how the reaction flow develops in this
low mass range of the $p$-process. This in particular will address
the question of feeding the $^{92,94}$Mo and $^{96,98}$Ru $p$-nuclei
which remain underproduced in present $p$-process nucleosynthesis
simulations \cite{ar03,rhhw02}.

\section{Investigated reactions}
\label{invesreac}

\begin{figure}
\resizebox{0.7\columnwidth}{!}{\rotatebox{270}{\includegraphics[clip=]{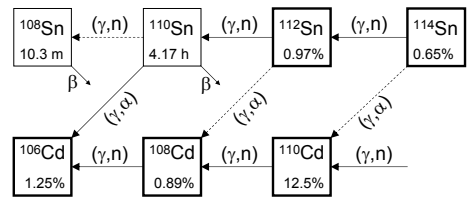}}}
\caption{The $p$-process reaction flow in the Cd-Sn region. For
simplicity, only even-even isotopes are shown, hence the
($\gamma$,n) arrow indicates two subsequent neutron emissions.
Stable isotopes are indicated by bold squares. The solid arrows
show the main reaction flow path while dashed arrows indicate
weaker branchings \cite{ra05}.} \label{fig_flow}
\end{figure}

The primary aim of the present study is to extend the existing
experimental database relevant to the $p$-process by measuring the
cross section of the \cdag reaction. Based on the Hauser Feshbach
predictions for the reaction rate, the $p$-process branching point
at which the ($\gamma,\alpha$) and ($\gamma$,p) reactions become
competitive with the ($\gamma$,n) process along the $Z$\,=\,50 isotopic
chain (Sn isotopes) is located at mass number region 110-112
\cite{ra05} (see Fig.\,\ref{fig_flow}). The ($\gamma,\alpha$) and
subsequent ($\gamma$,p) reactions on $^{110}$Sn and $^{112}$Sn
lead to the production of the $p$-nuclei $^{106}$Cd and $^{108}$Cd,
respectively ($^{108}$Cd has a slight contribution from the
$s$-process as well). The precise knowledge of these reaction rates
is essential to the reliable prediction of the $^{106}$Cd and
$^{108}$Cd abundances in $p$-process modeling. In the present work
the \cdag cross section is determined and the results are compared
with the prediction of statistical model calculations performed
with the \textsc{non-smoker} code \cite{NON-SMOKER} using different input
parameters such as optical model potentials and nuclear
level densities. In addition, the cross sections of the
\cdan reaction and the \cdap reaction below the ($\alpha$,n)
threshold have been measured and are compared with the \textsc{non-smoker}
predictions.

The peak of the Gamow window for the \cdag reaction at a
$p$-process temperature of $T_9$\,=\,3.0 is located at 7.21 MeV, its
width is about 4 MeV. 
The lowest energy reached in our experiments was 
$E_{\rm c.m.}$\,=\,7.56\,MeV (well within the Gamow window). The
measurements extended up to E$_{\rm c.m.}$\,=\,12.06\,MeV to probe the
reliability of the Hauser-Feshbach predictions over a wider energy
range.

\begin{figure}
\resizebox{0.8\columnwidth}{!}{\rotatebox{270}{\includegraphics{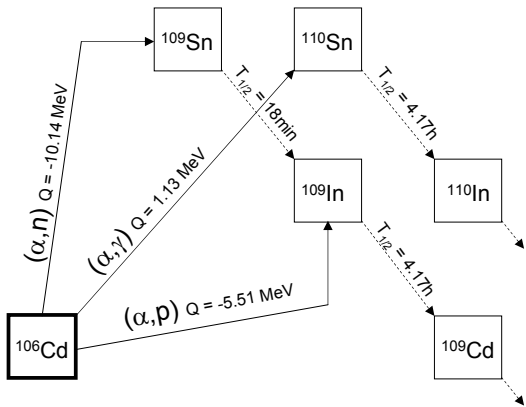}}}
\caption{The $\alpha$-induced reactions on $^{106}$Cd and the decay of the reaction products.}
\label{fig_decay}
\end{figure}

The reaction products of all three investigated reactions: $^{106}$Cd($\alpha,\gamma)^{110}$Sn,
\cdan and \cdap are radioactive. This makes it possible to
determine the cross sections using the activation technique. The
induced activity in a $^{106}$Cd target after bombarding with an
$\alpha$ beam can be measured off-line, and the above reaction
cross sections can be deduced from the measured $\gamma$-activity. The
reaction product of \cdap is the same as the daughter of
$^{109}$Sn from the \cdan reaction. Moreover, above the
($\alpha$,n) threshold (E$_{\alpha}$\,=\,10.53\,MeV), the
($\alpha$,n) channel becomes stronger than the ($\alpha$,p). Hence
the \cdap cross section is determined only below the ($\alpha$,n)
threshold (see Sec. \ref{sec_results}). The relevant part of the
chart of nuclides can be seen in Fig.\,\ref{fig_decay} where the
alpha-induced reactions and the decay of the reaction products are
shown. The decay parameters used for the analysis are summarized
in Table\,\ref{decaypar}.

\begin{table}
\caption{Decay parameters of the $^{106}$Cd\,+\,$\alpha$ reaction
products taken from the literature. For $^{110}$Sn
the half-life value from the literature is put in parentheses
since the recently determined precise half-life value from
\cite{gyu05} has been used for the analysis.}
\setlength{\extrarowheight}{0.1cm}
\begin{ruledtabular}
\begin{tabular}{ccccc}
\parbox[t]{1.5cm}{\centering{Product \\ nucleus}} &
\parbox[t]{1.5cm}{\centering{Half-life [hour]}} &
\parbox[t]{1.5cm}{\centering{Gamma \\energy [keV]}} &
\parbox[t]{1.5cm}{\centering{Relative \\ $\gamma$-intensity \\ per decay [\%]}} &
\parbox[t]{1.5cm}{\centering{Ref.}} \\
\hline
$^{110}$Sn &  4.173 $\pm$ 0.023 & 280.5 & 97 & \cite{gyu05} \\
& (4.11\,$\pm$\,0.1) & & & \cite{NDS89} \\ \\
$^{109}$Sn & 18.0 $\pm$ 0.2 min. &  1099.2 & 30.1 $\pm$ 3.0  & \cite{NDS06} \\
             &                 & 1321.3 & 11.9 $\pm$ 1.4 &  \\
&&&&\\
$^{109}$In & 4.167 $\pm$ 0.018 & 203.5  & 73.5 $\pm$ 0.5 & \cite{NDS06} \\
\end{tabular} \label{decaypar}
\end{ruledtabular}
\end{table}

\section{Experimental procedure}

In order to increase the reliability of the experimental results
and to find any hidden systematic error, the experiments have been
carried out independently in ATOMKI, Debrecen, Hungary and at
the University of Notre Dame, Indiana, USA. In the following
sections the experimental set-ups used in the two laboratories are
discussed.

\subsection{Experiments in ATOMKI}
\subsubsection{Target properties}

The targets were prepared by evaporating highly enriched
(96.47\,\%) $^{106}$Cd onto thin (d\,=\,3\,$\mu$m) Al foil. The Cd
powder was evaporated from a Mo crucible heated by electron
bombardment. The Al foil was placed 5\,cm above the crucible in a
holder defining a circular spot with a diameter of 12 mm on the
foil for Cd deposition. This procedure made it possible to determine
the target thickness by weighing. The weight of the Al foil was
measured before and after the evaporation with a precision better
than 5\,$\mu$g and from the difference the $^{106}$Cd number
density could be determined. Altogether 5 enriched targets were
prepared with thicknesses varying between 100 and
600\,$\mu$g/cm$^2$.

The thickness of the Al foil ensures that the heavy reaction
products are stopped in the backing. At the highest
$\alpha$-bombarding energy of 12.5\,MeV the energy of the
$^{110}$Sn recoil is 450\,keV and hence its range in Al is roughly
0.17\,$\mu$m, much smaller than the foil thickness.


\subsubsection{Activations}

\begin{figure}
\resizebox{\columnwidth}{!}{\rotatebox{270}{\includegraphics{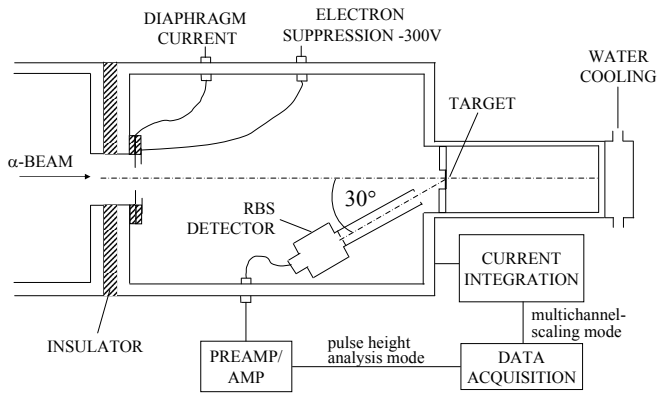}}}
\caption{Schematic view of the ATOMKI target chamber.}
\label{chamber}
\end{figure}

The activations have been performed at the MGC cyclotron at
ATOMKI. The energy range from E$_\alpha$=\,8.5 to 12.5\,MeV was
covered in 10 steps. The schematic view of the target chamber can
be seen in Fig.\ \ref{chamber}. After the last beam defining
aperture, the whole chamber served as a Faraday-cup to collect the
accumulated charge. A secondary electron suppression voltage of
$-300$ V was applied at the entrance of the chamber. Each
irradiation lasted about 10 hours and the beam current was
restricted to 500\,enA in order to avoid target deterioration. The
current was kept as stable as possible but to follow the changes
the current integrator counts were recorded in multichannel
scaling mode, stepping the channel in every minute. This recorded
current integrator spectrum was then used for the analysis solving
the differential equation of the population and decay of the
reaction products numerically.

A surface barrier detector was built into the chamber at
$\Theta$=150$^\circ$ relative to the beam direction to detect the
backscattered $\alpha$ particles and to monitor the target
stability this way. The RBS spectra were taken continuously and
the number of counts in the Cd peak was checked regularly during
the irradiation. Having the beam current restricted to 500\,enA, no
target deterioration was found within the precision of the RBS
measurement i.e. of the order of 1\%. Weighing the target 
foils after irradiation also confirmed this.

The beam stop was placed 10 cm behind the target from where no
backscattered particles could reach the surface barrier detector.
The beam stop was directly water cooled.

Because of an energy gap of the cyclotron, it is not possible to
accelerate $\alpha$-beam in the energy range between
E$_\alpha$\,=\,9 and 10.8\,MeV (with the exception of a narrow
allowed window around 10\,MeV where limited $\alpha$-current is
possible). Therefore the energy points of E$_\alpha$\,=\,9.5 and 10.33\,MeV have
been measured with higher energy beam and energy degrader foils.
For energy degrader Al foil with 8.8\,$\mu$m thickness was used.
The thickness was determined with measuring the energy loss of
$\alpha$-particles from an $\alpha$-source when passing through
the foil. The 10.33\,MeV energy was reached from the beam energy
of 11.2\,MeV using one layer of degrader foil while for the
9.5\,MeV point two layers of degrader foil and 11.3\,MeV beam was
used. In order to test the reliability of the degrader foil
method, the reaction cross section at 11.6\,MeV was measured
directly as well as using 12.4\,MeV beam and one degrader foil.
The two measurement gave the same result (see Sec.
\ref{sec_results}).

The highest energy point (E$_{\rm c.m.}$\,=\,12.06\,MeV) has also been
measured using a Cd target with natural isotopic composition. The
results with enriched and natural targets are the same within the
error (Sec. \ref{sec_results}).

\subsubsection{Detection of the induced $\gamma$-radiation}

The $\gamma$ radiation following the $\beta$-decay of the produced
Sn and In isotopes was measured with a HPGe detector of 40\%
relative efficiency. The target was mounted in a holder at a
distance of 10\,cm from the end of the detector cap. The whole
system was shielded by 10~cm thick lead against laboratory
background.

The $\gamma$-spectra were taken for at least 10 hours and stored
regularly in order to follow the decay of the different reaction
products.

The absolute efficiency of the detector was measured with
calibrated $^{133}$Ba, $^{60}$Co and $^{152}$Eu sources in the
same geometry used for the measurement. At E$_\gamma$\,=\,280.5\,keV 
the photopeak efficiency is (0.811\,$\pm$\,0.057)\,\%.

\begin{figure}
\vspace{0.5cm}
\resizebox{\columnwidth}{!}{\rotatebox{270}{\includegraphics{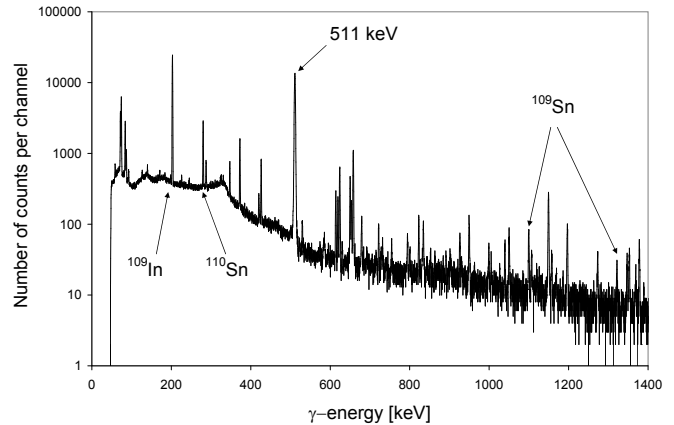}}}
\caption{Activation $\gamma$-spectrum after irradiating a target
with 12\,MeV $\alpha$-s. The $\gamma$ lines used for the analysis
and listed in Table \ref{decaypar} are indicated by arrows
together with the dominant 511\,keV annihilation line. The other
peaks are either from laboratory background or from the decay of
$^{109}$In (many weak transitions).} \label{gammaspec}
\end{figure}

Fig.\ \ref{gammaspec} shows an off-line $\gamma$-spectrum taken
after irradiation with 12~MeV $\alpha$-s in the first 1h counting interval.
The $\gamma$ lines used for the analysis are indicated by arrows.

Taking into account the detector efficiency and the relative intensity of the
emitted $\gamma$-rays, coincidence summing effects for all three reactions were
well below 1\% and were neglected.

\subsection{Experiments at Notre Dame}
\subsubsection{Target properties}

$^{106}$Cd targets used at Notre Dame were prepared by rolling
with nominal thicknesses of 2.3 mg/cm$^2$ and a $^{106}$Cd
enrichment of 86.4 \%.  The foils were mounted on Ta frames with
hole diameters of 12.5 mm. The actual thickness was determined
prior to the activations by Rutherford backscattering (RBS) to 2.1
$\pm$ 0.2 mg/cm$^2$. The targets were also monitored by RBS during
the activations (see below) and the thicknesses were again
verified after the conclusion of the experiment. Tests with
natural targets showed no deterioration of the targets when the
$\alpha$-beam currents were limited to $\le$ 300\,enA. Because the
Notre Dame experiment was designed to extend the ATOMKI data to
lower beam energies, the targets were not backed by a thin Al
layer to avoid the short lived $\gamma$-activity associated with
the Al activation. For this reason no waiting time was required
between the end of the activation and the counting; however, this
allows a small fraction of the heavy recoils to escape from the
target. At beam energies of $\le$10 MeV the target layer from
which recoils can escape is $\le$0.06 mg/cm$^2$ or less than 3
$\%$ of the target thickness. In addition, the cross section drops
significantly across the target thickness (factor of $\ge$ 5)
leading to an overall loss of activity of $\le$0.6 \%.

\subsubsection{Activations}

The activations at Notre Dame were carried out with the FN Tandem
Van de Graaff accelerator at beam energies between 7.0 and 12.0
MeV. However, no data could be obtained below 8 MeV because of
strong Compton background caused by a $\gamma$-line
(E$_\gamma$=373 keV) from the $^{40}$Ca($\alpha$,p)$^{43}$Sc
reaction. $^{40}$Ca is a common contaminant which has a Coulomb
barrier substantially lower than $^{106}$Cd and the half-life of $^{43}$Sc 
is similar to that of $^{110}$In. The cross section for this reaction only changes little over
the investigated energy range (see e.g. \cite{Howard74}) while the
$^{106}$Cd + $\alpha$ cross sections drop exponentially with beam
energy. The experimental setup of the target chamber was similar
to ATOMKI.  A collimator with a diameter of 5 mm defined the beam
spot. The isolated target chamber served as Faraday cup and a
suppression voltage of --300 V was applied to an isolated cathode
between collimator and chamber. In addition, a Si detector was
placed at 135$^\circ$ with respect to the beam direction to
monitor the target stability. The target was placed in a brass
holder which was air cooled and the beam was stopped in a thick
Carbon foil located directly behind the target. The digitized
charge and the energy signal of the Si detector were recorded
event-by-event together with the time of the event. Each
activation lasted 8 hours (approximately 2 half-lives) and the
beam current was kept below 300\,enA (see above).

\subsubsection{Detection of the induced $\gamma$-radiation}

The resulting $\gamma$-activity was measured with a pair of Ge
clover detectors which were mounted face to face. The detectors
were shielded by 5 cm of Pb against the room background and the 
distance between the detectors was 5 mm. 
The activated targets
were mounted in a holder which placed them at the center of the
detection system. The holder was made out of plastic and filled
out the whole space between the clover detectors except for the
space needed by the target. Each clover detector consists of 4
individual crystals with a relative efficiency of 20\,\%. 
The energies of the crystals were
recorded event-by-event together with the time of the event. In
addition a pulser signal was fed into the test input of one of the
Ge preamplifiers. This allowed to reconstruct the dead time as a
function of time. Each of the crystals were treated as an
independent Ge detector to reduce the problem of pile-up and
summing losses.

The absolute and relative $\gamma$-efficiencies of the detectors
were determined using calibrated $^{54}$Mn, $^{60}$Co, and
$^{133}$Ba sources as well as an uncalibrated $^{152}$Eu source.
The activity of the $^{152}$Eu source was determined relative to
the $^{54}$Mn and $^{60}$Co sources. The efficiency was determined
by two independent methods. In the standard method the known decay
branchings and activities of the sources were used to determine
the efficiency. However, because of the high counting efficiency
of the detector system, summing correction has to be applied for the
multiple line sources. These were taken from Ref~\cite{Shima}. 
The second method makes use of the high counting
efficiency and granularity of the detection system. Choosing
selected $\gamma$-transitions which are in sequence the detector
efficiency can be determined independent of the source
strength from ratios of single- and coincidence events. In this method, 
the problem of summing correction is either absent or
greatly removed. Both methods agreed within the uncertainties. 
The off-line detection system has a peak peak efficiency of (22.3\,$\pm$\,0.5)\,\% 
for a $\gamma$-energy of 280.5 keV.

\section{Experimental results}
\label{sec_results}

\begin{table}
\caption{Experimental cross section and $S$ factor of the
\cdag reaction}

\begin{ruledtabular}
\setlength{\extrarowheight}{0.1cm}
\begin{tabular}{lr@{\hspace{0.15cm}$\pm$\hspace{-0.25cm}}lr@{\hspace{0.15cm}$\pm$\hspace{-0.25cm}}
lr@{\hspace{0.15cm}$\pm$\hspace{-0.25cm}}l} E$_{\rm beam}$ &
\multicolumn{2}{c}{\hspace{-0.2cm}E$_{\rm c.m.}^{\rm eff}$} &
\multicolumn{2}{c}{\hspace{-0.4cm}Cross section} &
 \multicolumn{2}{c}{\hspace{-0.5cm}$S$ factor} \\
{[MeV]} & \multicolumn{2}{c}{\hspace{-0.2cm}[MeV]} & \multicolumn{2}{c}{\hspace{-0.4cm}[$\mu$barn]} &
\multicolumn{2}{c}{\hspace{-0.5cm}[10$^{21}$ MeV\,b]} \\
\hline
\multicolumn{7}{c}{ATOMKI results}\\
8.500 & 8.123 & 0.029 & 0.85 & 0.37 & 180 & 78 \\
9.008 & 8.632 & 0.026 & 4.87 & 0.55 & 155 & 18 \\
11.300\footnote{measured with energy degrader foil} & 9.108 & 0.049 & 22.8 & 2.9 & 143 & 18 \\
10.000 & 9.599 & 0.030 & 79.1 & 8.2 & 105 & 11 \\
11.200\footnotemark[\value{mpfootnote}] & 9.909 & 0.036 & 147 & 15 & 78.4 & 8.1 \\
10.800 & 10.371 & 0.033 & 234 & 24 & 34.3 & 3.5 \\
11.200 & 10.775 & 0.033 & 298 & 31 & 15.1 & 1.6 \\
11.600 & 11.167 & 0.034 & 507 & 53 & 9.8 & 1.0 \\
12.400\footnotemark[\value{mpfootnote}] & 11.167 & 0.037 & 471 & 49 & 9.08 & 0.94 \\
11.998 & 11.544 & 0.035 & 601 & 62 & 4.78 & 0.50 \\
12.523\footnote{measured with natural Cd target} & 12.050 & 0.036 & 1270 & 150 & 3.28 & 0.39 \\
12.523 & 12.057 & 0.036 & 1280 & 133 & 3.26 & 0.34 \\
\multicolumn{7}{c}{Notre Dame results} \\
8.000 & 7.566 & 0.010 & 0.078 & 0.014 & 164 & 29 \\
8.500 & 8.040 & 0.010 & 0.480 & 0.048 & 141 & 14 \\
9.000 & 8.513 & 0.011 & 2.59 & 0.26 & 126 & 13 \\
9.500 & 8.992 & 0.012 & 11.8 & 1.2 & 108 &  11 \\
10.000 & 9.466 & 0.012 & 46.4 & 4.6 & 92.7 & 9.3 \\
10.000 & 9.470 & 0.012 & 48.3 & 4.8 & 95.3 & 9.5 \\
10.000\footnote{measured with ATOMKI target} & 9.599 & 0.012 & 75.1 & 5.4  & 99.9 & 7.7 \\
11.000 & 10.429 & 0.014 & 244 & 124 & 30.6 & 3.6 \\
11.500 & 10.909 & 0.014 & 434 & 43 & 15.8 & 1.6 \\
12.000 & 11.385 & 0.015 & 596 & 61 & 6.85 & 0.70 \\
\end{tabular} \label{tab_ag}
\end{ruledtabular}

\end{table}

\begin{table}
\caption{Experimental cross section and $S$ factor of the
\cdan reaction}

\begin{ruledtabular}
\setlength{\extrarowheight}{0.1cm}
\begin{tabular}{lr@{\hspace{0.15cm}$\pm$\hspace{-0.25cm}}lr@{\hspace{0.15cm}$\pm$\hspace{-0.25cm}}
lr@{\hspace{0.15cm}$\pm$\hspace{-0.25cm}}l} E$_{\rm beam}$ &
\multicolumn{2}{c}{\hspace{-0.2cm}E$_{\rm c.m.}^{\rm eff}$} &
\multicolumn{2}{c}{\hspace{-0.4cm}Cross section} &
 \multicolumn{2}{c}{\hspace{-0.5cm}$S$ factor} \\
{[MeV]} & \multicolumn{2}{c}{\hspace{-0.2cm}[MeV]} & \multicolumn{2}{c}{\hspace{-0.4cm}[$\mu$barn]} &
\multicolumn{2}{c}{\hspace{-0.5cm}[10$^{21}$ MeV\,b]} \\
\hline
10.800 & 10.371 & 0.033 & 423 & 74 & 62 & 11 \\
11.200 & 10.775 & 0.033 & 1420 & 240 & 72 & 12 \\
11.600 & 11.167 & 0.034 & 2470 & 400 & 48 & 8 \\
12.400\footnote{measured with energy degrader foil} & 11.167 & 0.037 & 2600 & 440 & 50 & 8 \\
11.998 & 11.544 & 0.035 & 4785 & 720 & 38.0 & 5.7 \\
12.523 & 12.057 & 0.036 & 14400 & 2100 & 36.7 & 5.4 \\
\end{tabular} \label{tab_an}
\end{ruledtabular}

\end{table}

\begin{table}
\caption{Experimental cross section and $S$ factor of the \cdap reaction}

\begin{ruledtabular}
\setlength{\extrarowheight}{0.1cm}
\begin{tabular}{lr@{\hspace{0.15cm}$\pm$\hspace{-0.25cm}}lr@{\hspace{0.15cm}$\pm$\hspace{-0.25cm}}
lr@{\hspace{0.15cm}$\pm$\hspace{-0.25cm}}l} E$_{\rm beam}$ &
\multicolumn{2}{c}{\hspace{-0.2cm}E$_{\rm c.m.}^{\rm eff}$} &
\multicolumn{2}{c}{\hspace{-0.4cm}Cross section} &
 \multicolumn{2}{c}{\hspace{-0.5cm}$S$ factor} \\
{[MeV]} & \multicolumn{2}{c}{\hspace{-0.2cm}[MeV]} & \multicolumn{2}{c}{\hspace{-0.4cm}[$\mu$barn]} &
\multicolumn{2}{c}{\hspace{-0.5cm}[10$^{21}$ MeV\,b]} \\
\hline
\multicolumn{7}{c}{ATOMKI results}\\
10.000 & 9.599 & 0.030 & 6.0 & 0.8 & 8.0 & 1.1 \\
11.200\footnote{measured with energy degrader foil} & 9.909 & 0.036 & 39.6 & 4.2 & 21.1 & 2.2 \\
\multicolumn{7}{c}{Notre Dame results} \\
9.500 & 8.992 & 0.012 & 0.24 & 0.04 & 2.23 & 0.35 \\
10.000 & 9.470 & 0.012 & 2.76 & 0.28 & 5.45 & 0.55 \\
10.000\footnote{measured with ATOMKI target} & 9.599 & 0.01 & 5.70 & 0.67  & 7.58 & 0.89 \\
\end{tabular} \label{tab_ap}
\end{ruledtabular}

\end{table}

The cross sections and $S$ factors for the reactions \cdag, \cdan
and \cdap are listed in
Tables \ref{tab_ag}-\ref{tab_ap}, respectively. The second column
shows the effective center-of-mass energies \cite{Rolfs&Rodney}
which accounts for the decrease of the cross section over the
target thickness. For the ATOMKI measurement, the quoted errors of
the energies include the energy loss in the targets calculated
with the \textsc{srim} code \cite{SRIM}, the energy stability of the cyclotron and the
energy straggling in the degrader foil where it was applied. For
the Notre Dame results, the energy error is determined by the
uncertainty in the calculation of the effective energy. The
results obtained in ATOMKI and in Notre Dame are listed
separately. For the \cdan reaction no results from the Notre Dame
experiment are listed. The complex $\gamma$-decay scheme
\cite{NDS06} as well lack of any dominant $\gamma$-line and the
close geometry of the Notre Dame counting system require
significant summing corrections leading to large uncertainties.
For this reason we abstain from quoting any Notre Dame results for
this reaction.

Both sets of data are in excellent agreement 
(see tables \ref{tab_ag} and \ref{tab_ap}). To test for
systematic uncertainties, the \cdag cross section was measured at
Notre Dame at the same beam energy (10 MeV) using the same target
as in ATOMKI. The values are in excellent agreement (see Table
\ref{tab_ag}). The error of the cross section ($S$ factor) values
is the quadratic sum of the following partial errors: efficiency
of the detector system ($\sim$7\%(ATOMKI), $\sim$2.3\%(Notre
Dame)), number of target atoms ($\sim$6\%,$\sim$9\%), current
measurement (3\%,$\sim$3\%), uncertainty of the level parameters
found in literature ($\leq$12\%), counting statistics (0.1 to
40\%).

One possible uncertainty is the decay branching ratio of
$^{110}$Sn. The compilation \cite{NDS89} lists a 100 \% $\gamma$-branching
ratio to the 345 keV level in $^{110}$In. However, this value is
based only on an unpublished PhD thesis from 1956 \cite{Mead}. The
resulting log ft value of 3.24 is unusually small when compared
with other $\beta$-transitions between $\nu$(1$g_{9/2}$) and
$\pi$(1g$_{9/2}$) in this mass region and some of the decay branches
might not have been observed \cite{NDS89}. Calculation of the
\cdag cross section from the mother ($^{110}$Sn) and from the
daughter ($^{110}$In) activities provide an indirect way to
determine this decay branching ratio. This has been done for
several activations at ATOMKI and the resulting cross sections are always 
in good agreement within the errors. The weighted average of the ratios 
of the two cross sections from the two analyses is 1.041\,$\pm$\,0.073. 
This confirms the decay branching of 100 \% for the $^{110}$Sn 
$\beta$-decay.

While the \cdag reaction has been successfully observed for all
measured energies, \cdan cross sections can only be measured in
the upper half of the investigated energy region where the
($\alpha$,n) channel is open. Because of the problems described in
Sec.~\ref{invesreac}, the \cdap cross section has been determined
only below the \cdan threshold. Moreover, at the three lowest
measured energies the \cdap cross section is so low that no
cross section value could have been derived. Therefore, the \cdap
cross section has been measured only at 4 energies. 

\section{Discussion}

\subsection{Comparison to theory}

\begin{figure}
\vspace{0.5cm}
\resizebox{\columnwidth}{!}{\includegraphics{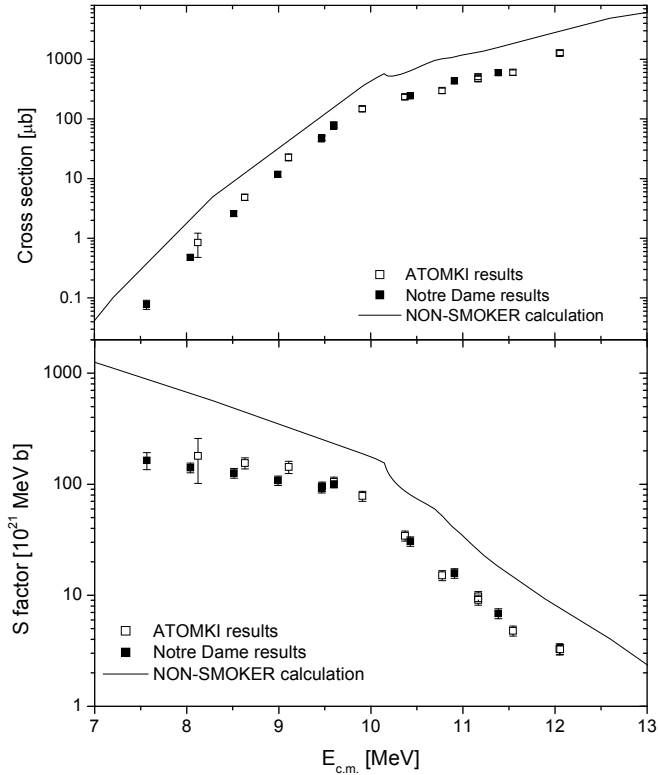}}
\caption{Cross section (upper panel) and $S$ factor of the \cdag reaction. 
ATOMKI data are represented with open
symbols, Notre Dame data with filled symbols. The line represents the results 
of the standard Hauser-Feshbach
calculation \cite{NON-SMOKERcs} (for details see text).} \label{ag_fig}
\end{figure}

\begin{figure}
\vspace{0.5cm}
\resizebox{\columnwidth}{!}{\includegraphics{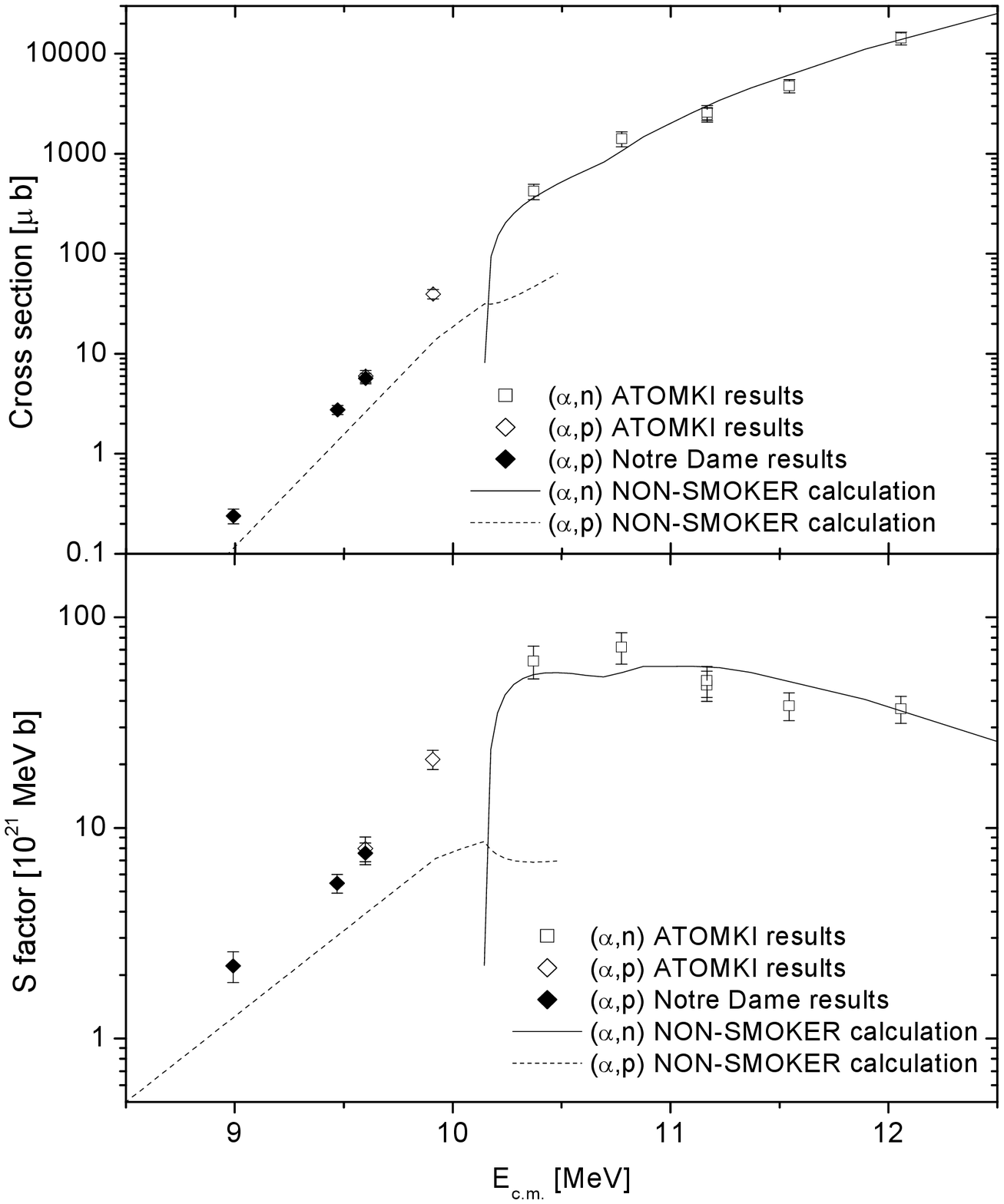}}
\caption{The same as Fig.~\ref{ag_fig} but for the \cdan and \cdap reactions.} 
\label{anp_fig}
\end{figure}

Figures \ref{ag_fig} and \ref{anp_fig} show a comparison of the experimental results to the Hauser-Feshbach statistical model cross sections \cite{NON-SMOKERcs} obtained with the standard settings of the \textsc{non-smoker}. There is excellent agreement for the ($\alpha$,n) reaction. The predicted cross sections are too low in the case of the ($\alpha$,p) reaction. Finally, the calculation yields cross sections which are too high by an almost constant factor of about 2.2 compared to
the ATOMKI data in the case of the ($\alpha$,$\gamma$) reaction.
While being at the same level of disagreement with the ($\alpha$,$\gamma$)
Notre Dame data above 9 MeV, the Notre Dame data for the three lowest
energies seem to indicate a different energy dependence than the theoretical
one. There is a factor of 5 disagreement between theory and experiment at the
lowest measured energy of 7.566 MeV.

\begin{figure}
\vspace{0.5cm}
\resizebox{\columnwidth}{!}{\includegraphics[angle=270]{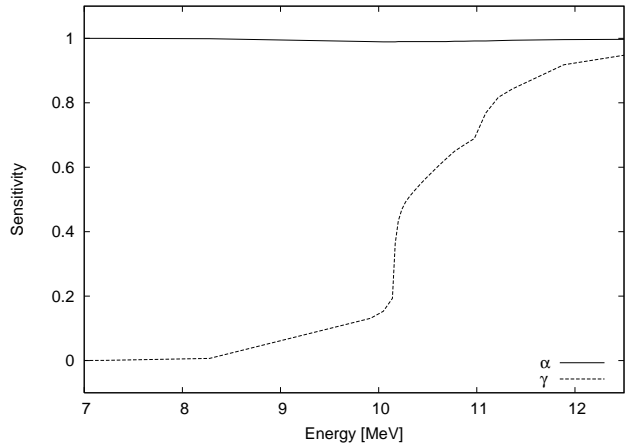}}
\caption{Sensitivity of the ($\alpha$,$\gamma$) cross section to a variation in the $\alpha$ and $\gamma$ widths, respectively. The sensitivity is given as function of $\alpha$ c.m. energy. It ranges from 0 (no change) to 1 (the cross section is changed by the same factor as the width).}
\label{sensi_fig}
\end{figure}

Since not only the capture reaction was measured but also the neutron and proton emission channels, interesting conclusions on the impacts of different inputs can be drawn. Usually, it is assumed that the $\gamma$ widths determine the cross sections in capture reactions because they are smaller than the particle widths. Inspection of the computed widths (directly derived from the calculated transmission coefficients; see \cite{koe04} for further details) shows that this is not the case here. Due to the small $Q$ value and the high Coulomb barrier the $\alpha$ widths are smaller or comparable. Only at the upper limit of the range of measured energies the cross section also becomes sensitive to the $\gamma$ width. This is shown in Fig.\ \ref{sensi_fig} where the sensitivity of the cross section (ranging between 0 for no sensitivity to 1 for full sensitivity) to variations in the $\alpha$ and $\gamma$ widths, respectively, is plotted. A similar comparison was performed for the other measured channels.
Our ($\alpha$,n) cross sections are sensitive to the $\alpha$ width and weakly dependent on the neutron width (except close to the threshold where the neutron width becomes smaller than the $\alpha$ width). The $^{106}$Cd($\alpha$,p) reaction is equally sensitive to $\alpha$ and proton widths.
All channels are quite independent of the nuclear level density because transitions to the low-lying states dominate and a number of these are explicitly included in the calculation (see \cite{NON-SMOKERcs} for a list of included states).

\begin{figure}
\vspace{0.5cm}
\resizebox{\columnwidth}{!}{\includegraphics[angle=270]{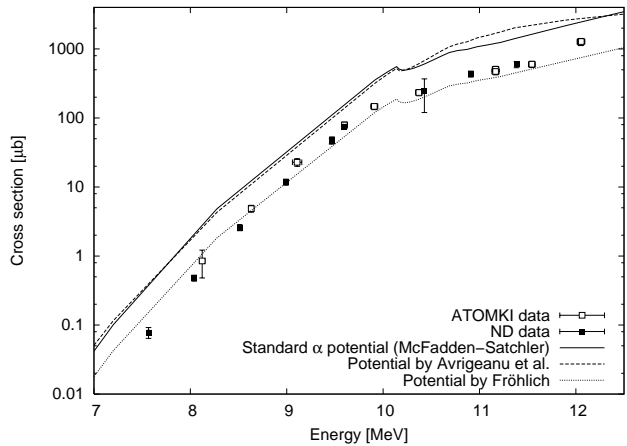}}
\caption{Experimental cross sections of the \cdag reaction compared to \textsc{non-smoker} predictions, using three different $\alpha$+nucleus potentials: by McFadden and Satchler \cite{McF66}, Avrigeanu {\it et al.}\ \cite{avri03}, and by Fr\"ohlich \cite{froh02,raufroh}.} 
\label{potentials_ag_fig}
\end{figure}

\begin{figure}
\vspace{0.5cm}
\resizebox{\columnwidth}{!}{\includegraphics{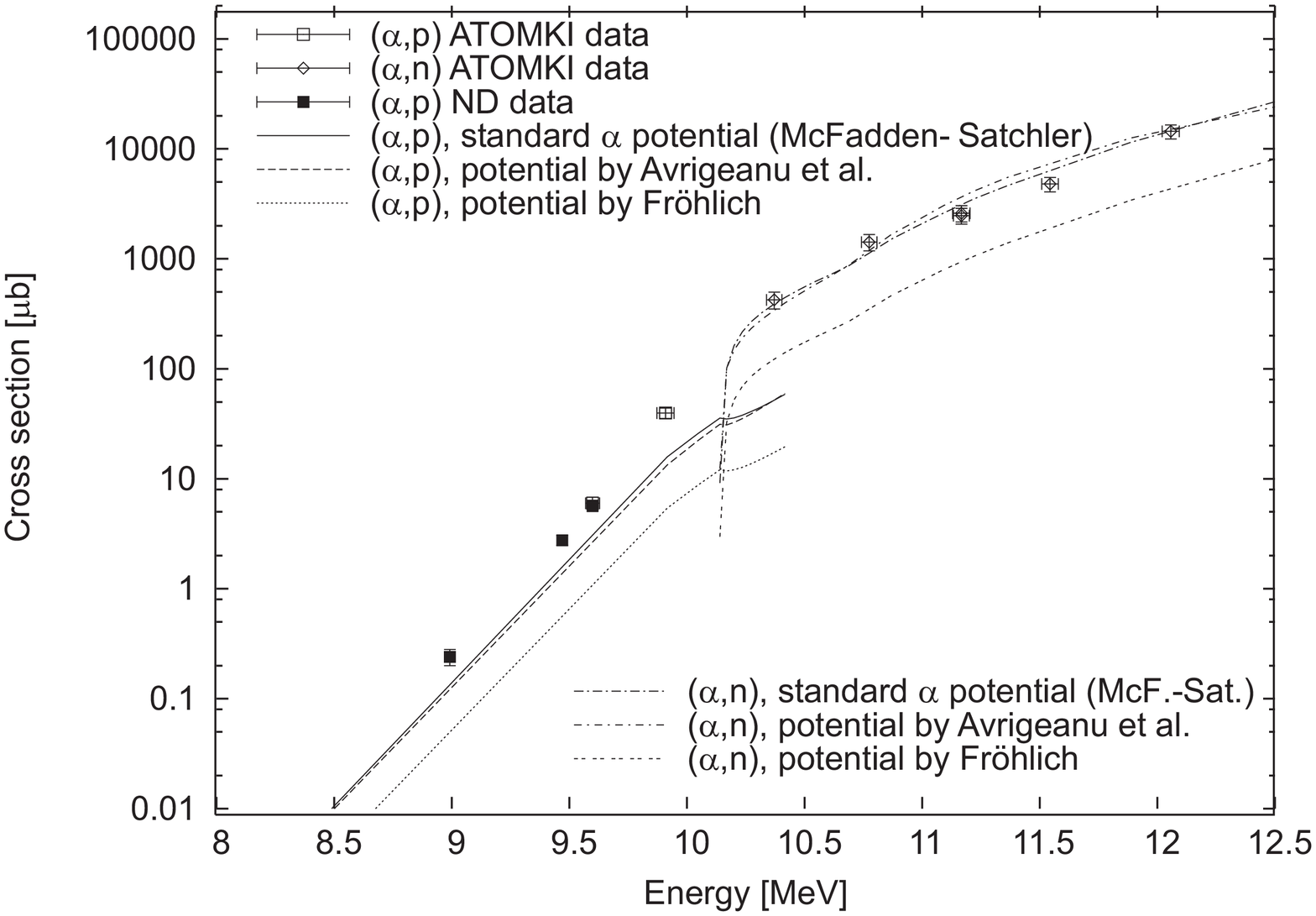}}
\caption{The same as Fig.~\ref{potentials_ag_fig} but for the \cdan and \cdap reactions.} 
\label{potentials_anp_fig}
\end{figure}

The standard predictions make use of the $\alpha$+nucleus optical potential by \cite{McF66}. Although this potential can reproduce scattering data over a large range of masses, it has been found to be problematic in describing $\alpha$ capture and emission at low energies (see, e.g., \cite{so98,koe04}). In Figs. \ref{potentials_ag_fig} and \ref{potentials_anp_fig} cross sections obtained with two different (more recent) $\alpha$+nucleus potentials are shown. Using the potential of \cite{avri03}, which was fitted on scattering data across a wide range of energies and masses, we obtain values not much different from those resulting from the use of the standard potential. Using the potential of \cite{froh02,raufroh}, which was fitted to (n,$\alpha$) and
($\alpha$,$\gamma$) reaction data around $A\simeq 145$, the capture cross sections are reduced by more than a factor of two and thus better agreement is found for the ($\alpha$,$\gamma$) reaction. However, due to the sensitivity of the ($\alpha$,p) and ($\alpha$,n) channels to the optical $\alpha$ potential, their cross sections are also reduced which removes the previously good agreement with the ($\alpha$,n) data and worsens the case of the ($\alpha$,p) reaction. It is interesting to note that recently the authors of Ref.\ \cite{avri03} have pointed out a possible difference between optical potentials derived from scattering data and such derived from reaction data. They conclude that optical potentials derived from scattering may have to be modified before applying them to reactions \cite{avri06}.

Figure \ref{sensi_fig} clearly shows the high sensitivity of the capture reaction to the $\alpha$ potential. This leads to the conclusion that indeed the $\alpha$ potential is the source of the disagreement with the experiment and that the capture reaction is best described with the potential of \cite{froh02,raufroh},
even though it appears to have a similar energy dependence as the standard potential and thus may still overestimate the cross sections at energies below the measured ones.
Fixing the $\alpha$ potential through the ($\alpha$,$\gamma$) reaction, one could assume that the disagreement between calculation and data for the ($\alpha$,p) channel has to arise from the proton optical potential. This conclusion appears puzzling when finally comparing with the ($\alpha$,n) reaction. For the latter we found only weak sensitivity to the neutron widths and compensating for the disagreement by changing the neutron width would require large modifications of the neutron optical potential. Since good agreement was found for neutron capture in this mass region \cite{bao}, it seems far-fetched to allow such a large modification. A possible explanation might be the fact that both ($\alpha$,p) and ($\alpha$,n) have a large negative $Q$ value and that therefore only few, low-energy neutron and proton transitions contribute. These will be very sensitive to the level scheme of low-lying levels. The spin assignments to these levels still bear considerable uncertainty even in the latest compilation \cite{NDS06} and the level schemes might still not be complete. This could explain why the neutron and proton emission channels are underestimated in the calculation.

\subsection{Astrophysical implications}

The standard statistical model rates of \cite{NON-SMOKERrates}, utilizing the potential of \cite{McF66}, are widely used in stellar models. For instance, with those rates the production of $p$-nuclei in the $p$-process in massive stars was studied in \cite{rhhw02}. Details on the branchings in the $p$-process path for the usual temperature range $2\leq T_9\leq 3$ can be found in \cite{ra05}. In Table I of Ref.\ \cite{ra05}, it can immediately be seen what changes are brought about by switching from the $\alpha$ potential of \cite{McF66} to the potential of \cite{froh02,raufroh} which better describes our capture data. The branching at $T_9=3$ remains unchanged (a ($\gamma$,p) branching at $^{110}$Sn) because photon-induced proton emission is by far faster than $\alpha$ emission. At $T_9=2.5$, proton and $\alpha$ emission start to compete when using the standard rates. However, with the lower rates obtained with the potential of \cite{froh02,raufroh}, $\alpha$ emission is still suppressed and proton emission is dominating. At $T_9=2.0$ the standard rates predict a ($\gamma$,$\alpha$) branching already at $^{112}$Sn. With the new potential the branching is shifted back to $^{110}$Sn and becomes a combined ($\gamma$,p)+($\gamma$,$\alpha$) branching. Considering that
the energy dependence at low energies may not be well reproduced and that the
actual cross section may be even lower, as indicated by the trend seen in the
low energy Notre Dame data, it is conceivable that the $\alpha$ branching at
$T_9=2.0$ is even further suppressed and may become negligible compared to
the $^{110}$Sn($\gamma$,p) branching.

The modification of the branching in the Sn chain by itself will not lead to large changes in the description of the $p$-process. However, our results show that the treatment of the optical $\alpha$ potential at astrophysically relevant energies will have to be improved. They also seem to suggest that the branchings obtained with Rate Set C of \cite{ra05} may be more accurate than those obtained with the standard Rate Set A concerning the $\alpha$ branchings.

\section{Summary}

The cross sections of three $\alpha$-induced reactions on $^{106}$Cd have been measured using the activation technique. The \cdag cross section was determined in the energy range between E$_{\rm c.m.}$\,=\,7.56 and 12.06\,MeV. Within this energy range, the particle emitting \cdan and \cdap cross sections have been measured above and below the ($\alpha$,n) threshold, respectively. 

For all three investigated reactions, the experimental results were compared with the cross sections calculated using the \textsc{non-smoker} statistical model code. The standard settings of the \textsc{non-smoker} code provided an excellent prediction for the ($\alpha$,n) cross section while the calculated cross sections were too low and too high for the ($\alpha$,p) and ($\alpha,\gamma$) channels, respectively. 
The sensitivity of the predictions to the input parameters was also examined.
It was found that good agreement with the experiment can be obtained
for the ($\alpha,\gamma$) channel by modifying the $\alpha$
optical potential. The same potential, however, simultaneously
results in a worse reproduction of the experimental results in
the ($\alpha$,n) and ($\alpha$,p) channels. The calculations for 
these channels, in turn, appear to have problems in other nuclear properties,
the proton and neutron optical potentials or, most likely, the uncertain
spin and parity assignments of excited states in the exit channels.

The value of the ($\alpha$,$\gamma$) reaction cross section and rate
influences the branching points in the $p$ process path. The impact of
the new experimental cross section, which is lower than previously predicted,
has been examined for the Sn isotopic chain.
The result underlines the importance of the experimental investigation
of $\alpha$-induced reactions in the mass and energy region relevant to the
astrophysical $p$ process.

\begin{acknowledgments}

We like to thank J. Greene from Argonne National Laboratory for
the preparation of the targets used at Notre Dame. The assistance
 of C. Ugalde, E. Strandberg, A. Couture, J. Couture, and E.
Stech during the course of the experiment is highly appreciated.
This work was supported by OTKA (Grant Nos. T042733, T049245,
F043408 and D048283), by the NSF-Grant PHY01-40324, MTA-OTKA-NSF grant 93/049901, 
the Swiss NSF 
(grants 2024-067428.01, 2000-105328), the Scientific and Technical 
Research Council of Turkey (TUBITAK) - Grant TBAG-U/111
(104T2467), and through the
Joint Institute of Nuclear Astrophysics (www.JINAweb.org) NSF-PFC
grant PHY02-16783. Gy.~Gy. and Zs.~F. acknowledge support from the Bolyai
grant.
\end{acknowledgments}

\end{document}